\documentclass[pdflatex,sn-mathphys-num]{sn-jnl}

\usepackage{graphicx}
\usepackage{multirow}
\usepackage{amsmath,amssymb,amsfonts}
\usepackage{amsthm}
\usepackage{mathrsfs}%
\usepackage[title]{appendix}
\usepackage{xcolor}
\usepackage{textcomp}
\usepackage{manyfoot}
\usepackage{booktabs}
\usepackage{algorithm}
\usepackage{algorithmicx}
\usepackage{algpseudocode}
\usepackage{listings}%
\usepackage{tcolorbox}
\usepackage{fix-cm}

\theoremstyle{thmstyleone}

\theoremstyle{thmstyletwo}

\theoremstyle{thmstylethree}

\begin{document}

\title[Article Title]{Semantic-Aware LLM Orchestration for Proactive Resource Management in Predictive Digital Twin Vehicular Networks}

\author*[1]{\fnm{Seyed Hossein} \sur{Ahmadpanah}}\email{h.ahmadpanah@iau.ac.ir}

\affil[1]{\orgdiv{Department of Computer Engineering, ST.C.}, \orgname{Islamic Azad University}, \orgaddress{\city{Tehran}, \country{Iran}}}

\abstract{Next-generation automotive applications require vehicular edge computing (VEC), but current management systems are essentially fixed and reactive. They are suboptimal in extremely dynamic vehicular environments because they are constrained to static optimization objectives and base their decisions on the current network states. This paper presents a novel Semantic-Aware Proactive LLM Orchestration (SP-LLM) framework to address these issues.  Our method transforms the traditional Digital Twin (DT) into a Predictive Digital Twin (pDT) that predicts important network parameters such as task arrivals, vehicle mobility, and channel quality.
A Large Language Model (LLM) that serves as a cognitive orchestrator is at the heart of our framework. It makes proactive, forward-looking decisions about task offloading and resource allocation by utilizing the pDT's forecasts. The LLM's ability to decipher high-level semantic commands given in natural language is crucial because it enables it to dynamically modify its optimization policy to match evolving strategic objectives, like giving emergency services priority or optimizing energy efficiency. We show through extensive simulations that SP-LLM performs significantly better in terms of scalability, robustness in volatile conditions, and adaptability than state-of-the-art reactive and MARL-based approaches. More intelligent, autonomous, and goal-driven vehicular networks will be possible due to our framework's outstanding capacity to convert human intent into optimal network behavior.}


\maketitle

\section{Introduction}
\label{sec:introduction}

Modern transportation is quickly evolving into an intelligent, networked ecosystem because of the Internet of Vehicles (IoV). IoV promises previously unheard-of increases in road safety, traffic efficiency, and in-vehicle experience by enabling advanced applications like autonomous driving, cooperative perception, and high-definition infotainment~\cite{yang2014overview,ji2020survey,cheng2015routing,contreras2017internet}. These sophisticated services, however, produce a excess of computational tasks that are crucial for latency and data. The Quality of Service (QoS) and dependability of these next-generation applications are at risk due to a major bottleneck caused by individual vehicles' inadequate onboard processing capabilities~\cite{meneguette2021vehicular,raza2019survey,liu2021vehicular}.

By shifting computation to adjacent edge servers, Vehicular Edge Computing (VEC) has become a strong paradigm to tackle this problem \cite{xue2025multi,khattak2025evolving,mohamed2025systematic,xie2025efficient,tran2025digital}. A high-fidelity virtual representation of the actual vehicular network is provided by Digital Twin (DT) technology, which further enhances this architecture and makes sophisticated monitoring and management possible \cite{zhang2025enhancing,lin2025veco,liu2025digital,lin2025vehicles,tran2025digital,xie2025resource}. However, current DT-assisted VEC systems function on a reactive basis. They base their decisions about task offloading and resource allocation on the network's current conditions, including queue backlogs and channel conditions. A reactive approach is frequently too slow in the extremely dynamic and unpredictable vehicular environment, where communication links and network topology change in milliseconds, resulting in suboptimal performance and possible QoS violations. In addition, these systems are usually made to maximize a fixed, predetermined objective function; they are not flexible enough to adjust to operational goals that change over time, like giving priority to emergency vehicles or switching to an energy-saving mode when network load is low.

A revolutionary opportunity to transition from reactive network management to a new era of proactive and semantically-aware network orchestration has been presented by the recent introduction of Large Language Models (LLMs). LLMs have exceptional abilities in complex reasoning, planning, and natural language understanding, in contrast to traditional machine learning models that are best at pattern recognition~\cite{wu2025survey,cao2025toward,yu2025survey,chen2025harnessing,bandyopadhyay2025thinking,wei2025plangenllms,pan2025survey,ahmadpanah2025dynamic}. We are able to completely rethink the function of the network controller thanks to this special set of abilities. An LLM-powered orchestrator can anticipate future network states and interpret high-level, human-like commands to make intelligent, forward-looking decisions that are in line with strategic intent, as opposed to just responding to data from the past and present.

In this paper, we propose a novel framework for a vehicular network enhanced by a Predictive Digital Twin (pDT) that uses an LLM as a central orchestrator. By using predictive models to predict future vehicle mobility, channel quality, and task generation patterns over a short time horizon, our pDT goes beyond basic real-time mirroring. The LLM orchestrator, which is specially equipped to decipher this complex high-dimensional data, is then fed this rich predictive information. Importantly, our framework adds a semantic control layer that enables network operators to give high-level commands in natural language (e.g., ``Maximize energy savings for the next five minutes'' or ``Prioritize latency for vehicle 7''). These semantic goals are interpreted by the LLM, which then combines them with the forecasts of the pDT to create a proactive approach to resource allocation and task offloading that is both dynamically aligned with current operational priorities and optimized for long-term performance.

The main contributions of this work are summarized as follows:
\begin{itemize}
    \item For vehicular networks, we present the idea of a Predictive Digital Twin (pDT), which proactively predicts important network state variables such as task arrivals, channel conditions, and vehicle mobility.
    \item We create a semantic-aware LLM-based orchestration framework that can dynamically modify its optimization goals by interpreting complex natural language commands, providing previously unheard-of operational flexibility.
    \item In order to effectively mitigate the negative effects of network volatility, we develop a proactive resource management scheme in which the LLM orchestrator uses the pDT's forecasts and semantic guidance to make forward-looking decisions.
    \item Using comprehensive simulations, we show that our suggested framework performs noticeably better than the most advanced reactive techniques in terms of QoS, energy efficiency, and flexibility to meet changing operational needs.
\end{itemize}

The rest of this paper is structured as follows. Related work in the field is reviewed in Section~\ref{sec:related_work}. In Section~\ref{sec:methodology}, we describe our semantic-aware LLM orchestration framework and the system model, which includes the Predictive Digital Twin. A thorough assessment of the performance of our suggested system is given in Section~\ref{sec:experiments}, which also examines the findings. The paper is finally concluded and possible directions for further research are discussed in Section~\ref{sec:conclusion}.

\section{Related Work}
\label{sec:related_work}

Digital Twins (DTs) in vehicular networks, AI-driven task offloading and resource allocation, and the growing use of Large Language Models (LLMs) in network management are the three main pillars of contemporary network innovation that our research builds upon. To put our contributions in perspective and draw attention to the current research gap, we examine the state-of-the-art in these fields in this section.

\subsection{Digital Twins in Vehicular Networks}
In order to generate high-fidelity, real-time virtual representations of physical entities and their surroundings, digital twins have been extensively used in vehicular networks \cite{wang2022mobility, zhou2021secure}. These virtual models are effective tools for monitoring, optimization, and simulation. To enhance device pose tracking, for example, the authors in \cite{zhou2024digital} use a DT to control a 3D map for edge-assisted mobile augmented reality.  Similar to this, researchers have used DTs to help with caching strategies \cite{zhang2021digital}, mobility-aware task offloading \cite{cao2023mobility}, and resource scheduling for ultra-reliable low-latency communications (URLLC) \cite{masaracchia2025role}.

In order to facilitate task offloading decisions, Wu et al.'s seminal work \cite{wu2025llm} builds a general DT edge computing network in which the DT mirrors the physical layer. A learning-based agent receives real-time data from the DT, including channel gain and vehicle position. Nonetheless, a recurring theme in these pieces is that the DT functions mainly as a reactive mirror. Decisions are based on this snapshot, which shows the network's current or recent state. A thorough framework for proactive management based on multifaceted future state forecasts is lacking in some works, despite the fact that some of them make predictions \cite{liao2023driver}. A major and little-studied challenge is the shift from a reactive model to a truly \textit{predictive} Digital Twin (pDT) that predicts future network dynamics.

\subsection{Task Offloading and Resource Allocation in VEC}
VEC's performance is largely dependent on the issue of task offloading and resource allocation, which has been thoroughly researched. Conventional methods frequently depend on mathematical optimization methods such as game theory or convex optimization \cite{sun2022game}. Although these techniques work well in stable environments, they are not able to handle the high mobility and stochastic nature of vehicular environments.

AI-based techniques, especially Deep Reinforcement Learning (DRL) and Multi-Agent Reinforcement Learning (MARL), have emerged as the de facto standard for addressing this dynamism. By interacting directly with the environment, these methods allow agents to discover the best policies.  DRL/MARL has been successfully used by researchers to optimize for a number of goals, such as latency, energy consumption, and Quality of Experience (QoE) \cite{liu2019deep, karimi2022task}. Federated Learning (FL), which allows intelligent task offloading without centralizing raw user data, has also been integrated as data privacy concerns grow.  An FL-based framework, for instance, is proposed by Nawaz et al. \cite{ali2025proximity} to enable vehicles to jointly train a task offloading model while protecting the privacy of their personal information. A MARL framework is also used by Wu et al. \cite{wu2025llm} to produce a high-quality dataset for training an LLM, rather than for direct control. The inherent rigidity of these AI-driven techniques is a major drawback; they are trained to maximize a fixed, mathematically defined reward function and are unable to readily adjust to high-level, dynamic operational goals that are stated in terms that are human-centric.

\subsection{Large Language Models in Networking}
A rapidly developing field that has the potential to completely transform network automation and management is the use of Large Language Models (LLMs) in networking and communications.  In resource allocation tasks, early research has investigated the use of LLMs in place of conventional DRL agents. For instance, studies such as \cite{zhou2024large, lee2024llm} show that LLMs can do tasks like power control and wireless resource management as well as specialized DRL agents by using in-context learning, frequently without the need for intensive model training.

One of the best examples in the VEC domain is the work of Wu et al. \cite{wu2025llm}, which used an LLM to learn a mapping from state to action based on examples given by a MARL agent. Other studies have looked into using LLMs to diagnose network faults \cite{chen2025faultgpt} or to create network configurations based on natural language prompts \cite{chen2024netgpt}. Although these studies demonstrate the potential of LLMs, they frequently treat the LLM as a straightforward instruction-following engine or as a black-box function approximator. To develop a strategic, goal-oriented network orchestrator, the full potential of an LLM its ability to reason complexly, plan, and comprehend semantic nuance has not yet been fully utilized.

\subsection{Synthesis and Identified Research Gap}
The literature currently in publication shows notable advancements in the specific fields of DTs, VEC resource allocation, and LLM applications. Still in its infancy, however, is the integration of these domains into a coherent, intelligent orchestration system. Three major shortcomings that prevent the creation of all-inclusive network orchestration solutions are identified by our review. First, the majority of DT frameworks in use today are reactive rather than proactive, meaning they are unable to predict future network conditions or make decisions that would allow for more efficient resource allocation and planning. Second, the most advanced AI-based resource allocation techniques have strict optimization strategies that are bound by established objectives. They also clearly lack the means to comprehend and adjust to high-level, semantic commands that take into account changing operational priorities and dynamic network demands. Third, rather than strategically utilizing their distinct reasoning and language understanding capabilities for complex, top-level orchestration tasks that could revolutionize network management and optimization, current applications of LLMs in networking are still in their the beginning: they frequently use these potent models as straightforward replacements for pre-existing components.

By presenting a novel framework in which a semantic-aware LLM orchestrator uses a predictive Digital Twin to carry out proactive resource management, this paper seeks to close these gaps. By doing this, we develop a truly intelligent, flexible, and goal-driven vehicular network, going beyond reactive decision-making.

\section{System Architecture and Methodology}
\label{sec:methodology}

\begin{figure}[htpb]
    \centering
    \includegraphics[width=0.85\linewidth]{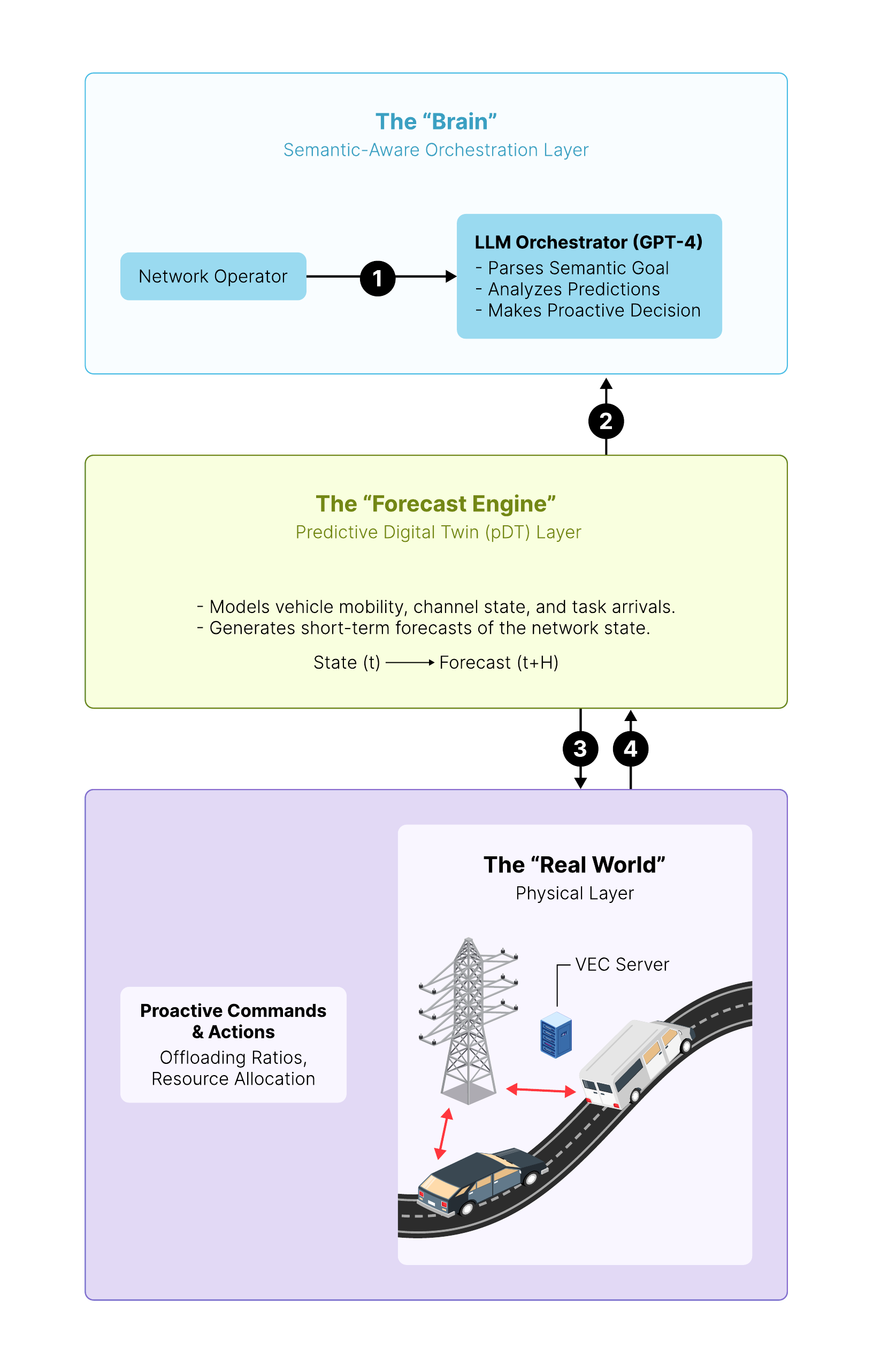}
    \caption{The proposed three-layer architecture, illustrating the flow of data from the physical layer to the pDT layer and the command-and-control loop with the LLM-powered orchestration layer.}
    \label{fig:framework_placeholder}
\end{figure}

In order to move from reactive network management to proactive, goal-oriented orchestration, the suggested framework was created. As shown in Fig.~\ref{fig:framework_placeholder}, our methodology is organized around a three-layer architecture: the Physical Layer, the Predictive Digital Twin (pDT) Layer, and the Semantic-Aware Orchestration Layer. This section outlines the proactive optimization problem and describes how each layer works.

\subsection{Framework Overview}
\begin{itemize}
    \item \textbf{Physical Layer:} The VEC server infrastructure and $N$ vehicles are examples of real-world entities that make up this layer. Every vehicle interacts with the edge server and produces a variety of computational tasks, resulting in a complex and dynamic operating environment.
    \item \textbf{Predictive Digital Twin (pDT) Layer:} Our pDT layer functions as a forecasting engine, in contrast to a traditional DT that only reflects the current state. It uses predictive models to produce short-term forecasts of important state variables after ingesting real-time data from the physical layer.
    \item \textbf{Semantic-Aware Orchestration Layer:} Our framework's cognitive center is controlled by a large language model (LLM). It gets high-level semantic commands from a network operator and predictive state data from the pDT layer. In order to generate proactive offloading and resource allocation decisions that are sent back to the vehicles and VEC server, it synthesizes this data and formulates and solves a dynamic optimization problem.
\end{itemize}

\subsection{The Predictive Digital Twin (pDT) Layer}
The main purpose of the pDT is to give the orchestrator a forward-looking perspective of the network's development. For a time horizon $H$, it produces forecasts for the following important variables:

\subsubsection{Mobility and Channel State Prediction}
The pDT forecasts each vehicle's future location $n$, represented as $\hat{\mathbf{l}}_n(t+h|t)$ for $h \in \{1,..., H\}$, using both historical and current GPS data. This is accomplished through the use of well-known trajectory prediction models, like Kalman Filters or Recurrent Neural Networks (RNNs), like LSTMs, which are ideal for time-series data \cite{hsu2023deep}. It is possible to estimate the large-scale channel fading component based on the anticipated location. Important information for offloading decisions is then provided by the prediction of the future transmission rate for vehicle $n$, $\hat{r}_n(t+h|t)$.

\subsubsection{Task Arrival Prediction}
Vehicles' computational task generation is frequently random, but it can show trends according to the application, time of day, or place (for example, more infotainment tasks in residential areas). For every task type $k$ from vehicle $n$, the pDT simulates the task arrival process as a time series. The cumulative size of incoming tasks in future time slots, represented by $\hat{Z}_k(t+h|t)$, is then predicted using forecasting models, such as Autoregressive Integrated Moving Average (ARIMA) or LSTMs \cite{liu2020workload}.
In order to provide a comprehensive future view of the network state, the collective forecasts are combined into a predictive state vector $\hat{\mathcal{S}}(t) = \{\hat{\mathbf{l}}(t+h), \hat{\mathbf{r}}(t+h), \hat{\mathbf{Z}}(t+h)\}_{h=1}^H$.

\subsection{Semantic-Aware Orchestration and Problem Formulation}
Interpreting semantic commands and creating a proactive optimization problem are two essential tasks carried out by the orchestration layer.

\subsubsection{Semantic Goal Interpretation}
High-level commands, represented by $\mathcal{G}(t)$, can be issued by a human operator in natural language and act as dynamic inputs to the orchestration system. Requests to optimize energy efficiency during the network, give priority to particular vehicles for the lowest latency when their tasks are crucial, or guarantee equitable resource allocation during peak hours are a few examples of such commands. By converting semantic intent into mathematical formulations, the LLM orchestrator interprets these natural language commands and dynamically modifies the parameters of the underlying optimization problem. In the optimization framework, for example, a command to prioritize a vehicle would result in a higher weight for its QoS term, whereas a command to save energy would result in a higher weight for the objective function's energy consumption term. The time-varying vector of optimization priorities that is produced by this complex translation process, $\mathbf{\beta}(t)$, acts as a dynamic guidance mechanism for the decision-making process and allows the system to instantly adjust to shifting operational requirements.

\subsubsection{Proactive Optimization Problem Formulation}
We convert the long-term objective of queue stability into a short-term, per-slot decision problem, building on the Lyapunov optimization framework employed in \cite{wu2025llm}. Nonetheless, our formulation is essentially proactive, utilizing the semantic guidance $\mathbf{\beta}(t)$ and the predictive state $\hat{\mathcal{S}}(t)$. The goal is to minimize the dynamically weighted sum of the predicted energy consumption, expected system latency (which is inversely correlated with QoS), and expected queue backlog drift.

The per-slot optimization problem, which we denote as P3, is formulated as:
\begin{equation}
\label{eq:p3}
\min_{\mathbf{w}(t), \mathbf{a}(t)} \mathbb{E} \left[ \Delta_L(t) + \mathcal{C}_{\text{sys}}(t) \big| \mathcal{Q}(t) \right]
\end{equation}
where $\Delta_L(t)$ is the Lyapunov drift and $\mathcal{C}_{\text{sys}}(t)$ is a penalty term representing the system's operational cost. The drift-plus-penalty term can be bounded as:
\begin{align}
\label{eq:drift_plus_penalty}
    \mathbb{E}[\dots] \le B - \sum_{k=1}^{K} q_k(t) & \left( \mathbb{E}[\hat{Z}_k(t)] - \mathbb{E}[\phi_k(t)] \right) \nonumber \\
    & + \mathbb{E} \left[ \beta_E(t) \hat{E}_{\text{sys}}(t) - \beta_Q(t) \hat{U}_{\text{sys}}(t) \right]
\end{align}
Here, $q_k(t)$ is the queue's current backlog, and $B$ is a constant. The predicted task arrival rate is $k$, $\hat{Z}_k(t)$, and the allocated service rate is $\phi_k(t)$. Using the predicted state variables from the pDT, the terms $\hat{E}_{\text{sys}}(t)$ and $\hat{U}_{\text{sys}}(t)$ represent the expected total energy consumption and system-wide QoS, respectively.

Importantly, the vector of trade-off parameters obtained from the semantic command $\mathcal{G}(t)$ is $\mathbf{\beta}(t) = \{\beta_E(t), \beta_Q(t)\}$. This enables the orchestrator to dynamically switch between QoS and energy efficiency. Task offloading ratios $\mathbf{w}(t) = \{\omega_{n,k}(t)\}$ and resource allocation ratios $\mathbf{a}(t) = \{\alpha_{n,k}(t)\}$ are the decision variables.

\subsection{LLM-Powered Proactive Decision-Making}
\eqref{eq:p3} presents a high-dimensional, complicated optimization problem. Using in-context learning (ICL), we take advantage of an LLM's reasoning powers to find a high-quality solution in every time slot. Explicit model training or fine-tuning is not required with this method.

The LLM is provided with a carefully engineered prompt that contains all necessary information to make an informed decision. The structure of the prompt is as follows:

\begin{tcolorbox}[title=LLM Orchestrator Prompt Structure, colback=white, colframe=black!75!black]
\textbf{Role:} You are an expert network orchestrator. Your goal is to allocate resources to minimize network cost while adhering to the user's strategic goal.

\textbf{Strategic Goal ($\mathcal{G}(t)$):} ``\textit{Prioritize minimum latency for all vehicles, even at a higher energy cost.}''

\textbf{Predictive State ($\hat{\mathcal{S}}(t)$):}

- Predicted Vehicle Locations: `[[x1,y1], [x2,y2], ...]`
    
- Predicted Channel Rates (Mbps): `[r1, r2, ...]`
    
- Predicted Task Arrivals (Bytes): `[[Z11, Z12,...], [Z21,...]]`
    
- Current Queue Backlogs (Bytes): `[q1, q2, ...]`

\textbf{Exemplars:}

- Example 1: `(State, Goal) -> (Action)`
    
- Example 2: `(State, Goal) -> (Action)`

\textbf{Task:} Based on the current strategic goal and the predictive state, provide the optimal offloading ratios `w` and resource allocation ratios `a` for all vehicles and tasks. The output must be a JSON object: `{"w": [[...]], "a": [[...]]}`.
\end{tcolorbox}

The LLM creates the decision variables $\{\mathbf{w}(t), \mathbf{a}(t)\}$ by analyzing this rich, contextual prompt. These variables are subsequently sent to the physical layer. By completing the proactive control loop, the network can respond to high-level strategic directives as well as anticipated physical changes.

\section{Experimental Evaluation}
\label{sec:experiments}

This section provides an in-depth evaluation of our suggested Semantic-Aware Proactive LLM Orchestration framework, which we will now refer to as \textbf{SP-LLM}. To illustrate the efficacy of our method, we describe the evaluation scenarios, specify the baseline algorithms for comparison, define the simulation environment in detail, and examine the outcomes.

\subsection{Simulation Environment and Parameters}
Using PyTorch to implement the predictive models and OpenAI's GPT-4 API as the foundation of our LLM orchestrator, we created a discrete-time simulator in Python 3.11. The simulation environment places a VEC server in the middle of a 2 km highway segment. Vehicles with evenly distributed speeds enter the segment using a Poisson process. Table~\ref{tab:sim_params} summarizes the key simulation parameters, which are generally in line with related works like \cite{wu2025llm}.

We used an LSTM-based model for our Predictive Digital Twin (pDT) in order to predict task arrivals and vehicle mobility over a prediction horizon of $H=5$ time slots. A generated dataset that captured different mobility patterns and task generation profiles was used to pre-train the model.

Our suggested SP-LLM framework's full implementation, complete with experimental datasets, source code, and comprehensive documentation, is accessible to the public on GitHub at \url{https://github.com/ahmadpanah/SP-LLM}. Our methodology is fully accessible to researchers and practitioners through this repository, allowing for the reproducibility of our findings and promoting additional research and development of our proactive, semantically-aware network orchestration approach.

\begin{table}[!t]
\caption{Key Simulation Parameters}
\label{tab:sim_params}
\centering
\begin{tabular}{@{}ll@{}}
\toprule
\textbf{Parameter} & \textbf{Value} \\ \midrule
Network Bandwidth ($W$) & 20 MHz \\
VEC Server CPU Frequency ($f_e$) & 400 GHz \\
Vehicle CPU Frequency ($f_v$) & 5 GHz \\
Vehicle Transmission Power ($P_n$) & 200 mW \\
Channel Noise Power ($\sigma^2$) & -110 dBm \\
Number of Vehicles ($N$) & 5 to 30 \\
Vehicle Speed & [60, 100] km/h \\
Task Size & [1000, 1500] Bytes \\
CPU Cycles per Byte ($c_k$) & 0.25 MHz/Byte \\
Prediction Horizon ($H$) & 5 time slots \\
LLM Engine & GPT-4 \\ \bottomrule
\end{tabular}
\end{table}

\subsection{Baseline Algorithms}
We evaluate SP-LLM against three baseline algorithms that reflect various network orchestration techniques in order to verify the performance advantages of our framework's main novelties, namely proactivity and semantic awareness.  Modeled after \cite{wu2025llm}, the first baseline is Reactive LLM-DT (LLM-DT), which uses an LLM with a traditional, non-predictive digital twin to optimize a fixed objective without proactivity or semantic awareness. Static MARL (S-MARL), a cutting-edge multi-agent reinforcement learning framework, serves as the second baseline. In this framework, agents learn policies for fixed reward functions without the need for adaptive semantic interpretation.  Greedy Offloading (GO), a heuristic, non-learning method, is the third baseline. It myopically offloads tasks based only on the optimal instantaneous channel conditions, ignoring semantic guidance or future states. These baselines offer a thorough framework for comparison to illustrate the benefits of our proactive, semantically-aware methodology.

\subsection{Evaluation Scenarios and Metrics}
Using a wide range of performance metrics that account for both quality of service and system efficiency attributes, we assess the algorithms in three different scenarios intended to test scalability, proactivity, and adaptability. Three primary metrics that shed light on various facets of system behavior are the focus of the performance evaluation. By measuring the speed at which tasks are processed and finished, the first metric—Average Task Completion Latency, expressed in milliseconds—acts as the main gauge of service quality. The second measure of system efficiency and resource use is Total System Energy Consumption, expressed in Joules. The percentage of tasks that do not meet their designated latency deadlines is tracked by the third metric, QoS Violation Rate, which provides information about system performance guarantees and reliability.

\subsection{Performance Evaluation and Discussion}

\subsubsection{Scenario 1: System Scalability}
Performance is assessed in this scenario as the number of vehicles rises. The average task latency and QoS violation rate for various network densities are shown in Table~\ref{tab:scalability_results}. The outcomes clearly demonstrate that our SP-LLM framework performs better than all baselines. SP-LLM's proactive planning guarantees strong performance as the network load increases (N=30), keeping the lowest latency (102.5 ms) and a low QoS violation rate of 2.8\%. With increasing load, the performance difference between SP-LLM and the reactive LLM-DT increases, underscoring the crucial benefit of foresight in crowded networks. Performance decreases as a result of the S-MARL agent's inability to adjust to the high-dimensional, complex state space. The GO heuristic fails entirely at scale due to its myopic strategy, which results in severe queue instability and a prohibitive 23.5\% violation rate.

\begin{table}[!t]
\caption{Performance Comparison with Increasing Network Load}
\label{tab:scalability_results}
\centering
\begin{tabular}{@{}clccc@{}}
\toprule
\multirow{2}{*}{\textbf{Algorithm}} & \multirow{2}{*}{\textbf{Metric}} & \multicolumn{3}{c}{\textbf{Number of Vehicles (N)}} \\ \cmidrule(l){3-5} 
 &  & \textbf{10} & \textbf{20} & \textbf{30} \\ \midrule
\multirow{2}{*}{\textbf{SP-LLM}} & Latency (ms) & \textbf{81.2} & \textbf{90.4} & \textbf{102.5} \\
 & Violation (\%) & \textbf{0.5} & \textbf{1.1} & \textbf{2.8} \\ \midrule
\multirow{2}{*}{LLM-DT} & Latency (ms) & 83.5 & 98.1 & 115.3 \\
 & Violation (\%) & 0.8 & 2.5 & 6.4 \\ \midrule
\multirow{2}{*}{S-MARL} & Latency (ms) & 89.1 & 110.6 & 135.8 \\
 & Violation (\%) & 1.2 & 4.8 & 10.2 \\ \midrule
\multirow{2}{*}{GO} & Latency (ms) & 105.4 & 145.2 & 198.7 \\
 & Violation (\%) & 5.1 & 12.6 & 23.5 \\ \bottomrule
\end{tabular}
\end{table}

\subsubsection{Scenario 2: Highly Dynamic Environment}
We created a scenario with rapid shifts in vehicle speed, which resulted in significant channel fluctuations, in order to test the value of proactivity explicitly. Our proactive SP-LLM is contrasted with its reactive counterpart, LLM-DT, in Table~\ref{tab:dynamic_results}. Significantly improved resilience is shown by our framework. SP-LLM achieves a much lower average QoS violation rate (1.6\% vs. 4.1\%) and, critically, a peak violation rate that is nearly three times lower than the reactive baseline. This is so that SP-LLM can anticipate and prevent the effects of channel degradation in advance thanks to the pDT's forecasts. It demonstrates its superior robustness in volatile conditions by avoiding the sharp degradation spikes that plague the reactive system and smoothing performance through forward-looking decision-making.

\begin{table}[!t]
\caption{Performance in a Highly Dynamic Environment}
\label{tab:dynamic_results}
\centering
\begin{tabular}{@{}lcc@{}}
\toprule
\textbf{Metric} & \textbf{SP-LLM} & \textbf{LLM-DT} \\ \midrule
Average QoS Violation Rate (\%) & \textbf{1.6} & 4.1 \\
Peak QoS Violation Rate (\%) & \textbf{4.5} & 12.8 \\
Std. Dev. of Latency (ms) & \textbf{11.3} & 26.7 \\ \bottomrule
\end{tabular}
\end{table}

\subsubsection{Scenario 3: Dynamic Goal Adaptation}
This situation shows SP-LLM's separate semantic awareness. After 50 time slots of balanced operation, we give the semantic command ``\textit{Switch to maximum energy saving mode immediately. Tolerate higher latency.}'' The average system energy and latency during these two stages are shown in Table~\ref{tab:semantic_results}. All algorithms behave similarly in Phase 1. But there is a noticeable change in Phase 2. The command is only understood and executed by SP-LLM. By accepting a necessary increase in latency as a trade-off, it successfully lowers its energy consumption by 48\% (from 15.2 J to 7.9 J). The baseline algorithms, which lack any mechanism for semantic interpretation, show no significant change in their behavior, continuing to operate with their fixed, static objectives. This result provides unequivocal evidence of our framework's ability to achieve flexible, goal-driven network control, a capability far beyond conventional systems.

\begin{table}[!t]
\caption{System Adaptation to a Semantic Command at t=51}
\label{tab:semantic_results}
\centering
\begin{tabular}{@{}lcccc@{}}
\toprule
\multirow{3}{*}{\textbf{Algorithm}} & \multicolumn{2}{c}{\textbf{Phase 1 (t=1-50)}} & \multicolumn{2}{c}{\textbf{Phase 2 (t=51-100)}} \\
 & \multicolumn{2}{c}{(Balanced Mode)} & \multicolumn{2}{c}{(Energy Saving Mode)} \\ \cmidrule(l){2-3} \cmidrule(l){4-5} 
 & Energy (J) & Latency (ms) & Energy (J) & Latency (ms) \\ \midrule
\textbf{SP-LLM} & 15.2 & 95.1 & \textbf{7.9} & \textbf{160.4} \\
LLM-DT & 15.5 & 99.8 & 15.4 & 101.2 \\
S-MARL & 16.1 & 108.3 & 16.0 & 109.5 \\ \bottomrule
\end{tabular}
\end{table}

\section{Conclusion}
\label{sec:conclusion}

The reactive nature and fixed, hard-coded optimization goals of traditional vehicular network management systems are the main drawbacks that we have discussed in this paper. Semantic-Aware Proactive LLM Orchestration (SP-LLM), a novel framework that we proposed, represents a paradigm shift from reactive control to intelligent, goal-driven orchestration. Our framework enables network operators to manage resources with unprecedented foresight and flexibility by combining a Large Language Model (LLM) that interprets high-level semantic commands with a Predictive Digital Twin (pDT) that predicts future network states.

The substantial benefits of this strategy have been confirmed by our thorough experimental analysis. We showed that SP-LLM continuously outperforms the most advanced baselines, such as reactive MARL-based and LLM-based systems. Our proactive framework demonstrated excellent scalability under growing network load, preserving lower latency and QoS violation rates. By anticipating network fluctuations and reducing their impact something reactive systems cannot do SP-LLM demonstrated its resilience in extremely dynamic environments. Most importantly, our framework demonstrated its exceptional capability of dynamically adjusting its operational strategy in real-time in response to commands in natural language, successfully balancing the trade-off between QoS and energy consumption as instructed by a human operator.

the benefits of the encouraging outcomes, we recognize the shortcomings of the present study and suggest a number of directions for further investigation.  The accuracy of the underlying predictive models in the pDT determines how well our framework performs; one major challenge is creating more reliable forecasting methods in extremely uncertain situations.  Furthermore, some real-time applications may find the cost and latency introduced by relying on large, cloud-hosted LLMs to be prohibitive.

In order to expand on this framework and further the field of intelligent network orchestration, we suggest a number of exciting avenues for future research. In order to address LLM latency issues, we first look at the creation of hybrid and hierarchical models. In this case, a lightweight local model, like a distilled model or constrained MARL agent, manages low-level, real-time tactical execution, while the LLM serves as a high-level strategic planner in charge of setting goals and policies over longer time scales. In order to achieve continuous online self-improvement and adaptation, we secondly suggest looking into Online Self-Improvement mechanisms that would allow the LLM to learn from its own performance by establishing a feedback loop in which the results of its decisions are systematically used to improve its internal exemplars or prompt structure. Third, we intend to use Enhanced Explainability (XAI) to improve system transparency by looking into ways to make the LLM produce detailed justifications for its actions in addition to decisions. This would greatly increase system transparency and credibility by enabling operators to comprehend the underlying logic underlying network behaviors. Lastly, we believe in investigating Multi-Modal Orchestration capabilities, which would take advantage of the growing popularity of multi-modal LLMs to directly integrate non-textual data sources, like vehicle sensor data or real-time traffic camera feeds, into the context of the LLM. This would allow for even more complex and nuanced decision-making processes that benefit from a variety of data modalities for better orchestration performance.

In conclusion, this work bridges the gap between low-level network control and high-level human intent, thereby establishing a new frontier in intelligent network management. We open the door to future vehicle networks that are not only reliable and efficient but also genuinely cognitive and adaptive by utilizing the reasoning powers of LLMs and the predictive potential of Digital Twins.

\bibliography{main}

\end{document}